\documentclass[fleqn,12pt]{SelfArx} 

\usepackage[english]{babel} 
\usepackage{tikz}
\usepackage{natbib}
\usepackage[normalem]{ulem}
\useunder{\uline}{\ul}{}
\definecolor{color1}{RGB}{0,0,90} 
\definecolor{color2}{RGB}{0,20,20} 

\usepackage{hyperref} 
\hypersetup{hidelinks,colorlinks,breaklinks=true,urlcolor=color2,citecolor=color1,linkcolor=color1,bookmarksopen=false,pdftitle={Title},pdfauthor={Author}}

\defcitealias{aaas90}{AAAS 1990}
\defcitealias{nrc07}{NRC 2007}
\defcitealias{nrc10}{NRC 2010}
\defcitealias{martin18}{Martin et al., 2018}
\defcitealias{reichart05}{Reichart et al., 2005}
\defcitealias{slater18}{Slater, 2018}

\JournalInfo{Robotic Telescopes, Student Research and Education (RTSRE) Proceedings} 
\Archive{ Conference Proceedings, San Diego, California, USA, Jul 23-27, 2018 \\Editor, E., Editor E., Eds. Vol. 1, No. 1, (2018) \\ ISSN XXXX-XXXX (online) / doi : doidoidoidoi / CC BY-NC-ND license \\ Peer Reviewed Article. rtsre.org/ojs} 
\PaperTitle{Factors Contributing to Attitudinal Gains in Introductory Astronomy Courses} 

\Authors{Adam S. Trotter\textsuperscript{1}, Daniel E. Reichart\textsuperscript{1}*, Aaron P. LaCluyz\'{e}\textsuperscript{2} \& Rachel Freed\textsuperscript{3}} 
\affiliation{\textsuperscript{1}\textit{Department of Physics \& Astronomy, University of North Carolina at Chapel Hill}} 
\affiliation{\textsuperscript{2}\textit{Department of Physics, Central Michigan University}} 
\affiliation{\textsuperscript{3}\textit{Sonoma State University}} 
\affiliation{*\textbf{Corresponding author}: reichart@physics.unc.edu} 
\affiliation{}
\affiliation{\textit{Received ?; revised ?; accepted ?}} 

\Keywords{} 
\newcommand{\keywordname}{Keywords} 

\Abstract{Most students do not enroll in introductory astronomy as part of their major; for many, it is the last science course they will ever take. Thus, it has great potential to shape students’ attitudes toward STEM fields for the rest of their life. We therefore argue that it is less important, when assessing the effectiveness of introductory astronomy courses, to explore traditional curricular learning gains than to explore the effects that various course components have on this attitude. We describe the results of our analysis of end-of-semester surveys returned by a total of 749 students in 2014-2015, at 10 institutions that employed at least part of the introductory astronomy lecture and lab curriculum we first implemented at the University of North Carolina at Chapel Hill in 2009. Surveys were designed to measure each student’s attitude, and to probe the correlation of attitude with their utilization of, and satisfaction with, various course components, along with other measures of their academic background and their self-assessed performance in the course. We find that students’ attitudes are significantly positively correlated with the grade they expect to receive, and with their rating of the course’s overall effectiveness. To a lesser degree, we find that students’ attitudes are positively correlated with their mathematical background, with whether they intend to major or pursue a career in STEM, and with their rating of the effectiveness of the instructor. We find that students’ attitudes are negatively correlated with the amount of work they perceived the course to involve, and, surprisingly, with the size and reputation of their home institution. We also find that, for the subsets of students who were exposed to them, students’ attitudes are positively correlated with their perception of the helpfulness of the lecture component of the course, and of telescope-based labs that utilized UNC-CH’s Skynet Robotic Telescope Network.
}

\begin{document}

\maketitle

\begin{tikzpicture}[remember picture,overlay]
   \node[anchor=north west,inner sep=10pt, xshift=1.5cm, yshift=-0.25cm] at (current page.north west)
              {\includegraphics[scale=2.0]{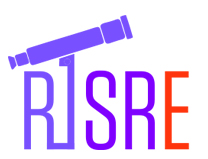}};
\end{tikzpicture}

\flushbottom 

\maketitle 


\thispagestyle{empty} 

\section*{Background}
Increasing interest and participation in STEM fields has been a major goal at the national level for many years, as the United States struggles to keep up globally with scientific and engineering pursuits \citepalias{aaas90, nrc07} while simultaneously declining in global rankings of science education \citep{kastberg16, provasnik16}. Meanwhile, little progress has been made on the front of increasing the number and quality of highly trained scientists and engineers in this country or of producing scientifically literate citizens \citep{alper16}.

Concurrently, as a potential solution to these issues, there have been dozens of attempts over the past two and a half decades to provide telescope access for education \citep{gomez17}, often under the presumption made by project personnel that, if the telescope is available and accessible, educators and students will inevitably use it for learning \citep{slater14}. In contrast, many of the programs developed over the past 25 years have not succeeded in their goals, with several even failing to launch after publication of their intended existence. 

Astronomy is often referred to as the ``Gateway Science'' \citepalias{nrc10}, with an estimated 240,000 students taking introductory astronomy or ``Intro Astro'' in the US, according to a 2012 survey by the American Institute of Physics \citep{mulvey14}.  It is often noted that Intro Astro is the last science class many students will ever take and is thus poised in an important position to promote scientific understanding and literacy for citizens as they leave the academic world and enter the workforce.

Skynet \citepalias{reichart05, martin18} has in large part solved the decades-old struggle to provide telescope learning experiences for students, particularly at large enrollment scales. Since its inception in 2004, Skynet has grown to one of the largest robotic telescope networks in the world, with nearly 30 optical telescopes ranging in size from 14 to 40 inches in diameter, a 20-meter radio telescope, and with several more telescopes soon to be added. These telescopes are all controlled through a web-based portal used by professional astronomers and students alike.  Approximately 50,000 students, from middle school through to senior undergraduate, have used Skynet to date. 

A few researchers have pointed out that  the value of remote telescope use in settings with large enrollments is unclear due to the current lack of risk-benefit analysis in the literature \citepalias[e.g.,][]{slater18}.  While much of the focus on astro101 has been on learning gains \citep[e.g.,][]{prather09, schlingman12, williamson16}, much less attention has been paid to attitudes towards science, and  astronomy in particular, in Intro Astro. This is due in part to a lack, until recently, of reliable and validated attitude assessment tools for astronomy \citep{bartlett18}, but also to the difficulties of curriculum design connecting expensive telescope resources to large enrollments \citep{slater07}.

In this paper, we explore the effects on students’ attitudes towards astronomy \citep{zeilik99}, based on responses to end-of-semester surveys of 749 Intro Astro students at 10 institutions between 2014 and 2015. These students undertook, in whole or in part, an introductory astronomy lecture and lab curriculum first implemented at University of North Carolina at Chapel Hill (UNC-CH). This is the first known exploration of students’ attitudes combining robotic telescopes and large enrollment Intro Astro courses.

\section*{Project Intro Astro} 

\addcontentsline{toc}{section}{Project Intro Astro} 


In 2009, we introduced a new introductory astronomy lecture and lab curriculum at UNC-CH. At most universities, introductory astronomy is taught as a two-semester sequence, but at UNC-CH it had always been taught in a single semester, which for the students was akin to drinking from a fire hose. In 2009, we split the old course into two new courses:
\\
\\
\noindent \textbf{ASTR 101:  The Solar System}\\ \textit{Celestial motions of Earth, the sun, the moon, and the planets; the nature of light; ground and space-based telescopes; comparative planetology; Earth and the moon; terrestrial and gas planets and their moons; dwarf planets, asteroids, and comets; planetary system formation; extrasolar planets; the search for extraterrestrial intelligence (SETI).}
\\
\\
\noindent \textbf{ASTR 102:  Stars, Galaxies, and Cosmology}\\ \textit{The sun; stellar observables; star birth, evolution, and death; novae and supernovae; white dwarfs, neutron stars, and black holes; Einstein’s theory of relativity; the Milky Way galaxy; normal galaxies, active galaxies, and quasars; dark matter and dark energy; cosmology; the early universe.}
\\

\noindent This created time to explore the material more thoroughly and more enjoyably, to introduce new material (e.g., a week of relativity in ASTR 102), and to introduce in-class demonstrations. Altogether, we developed over 50 in-class demonstrations, which we found to be particularly effective at conveying otherwise difficult concepts and at generating discussion, even in the largest classes.  We have now taught these courses successfully to as few as approximately 10 students and to as many as approximately 400 students, where so far success has been measured by end-of-course evaluations that are among the highest in our department, as well as by rapidly growing introductory astronomy enrollment. 

The centerpiece of our new introductory astronomy curriculum has been the modernization of our introductory astronomy laboratory course, ASTR 101L. For decades, ASTR 101L made use of the theater of the Morehead Planetarium and Science Center on the UNC-CH campus, for five day labs and small telescopes on our campus observing decks for five night labs.  However, both sets of labs were problematic. Measurements within the planetarium chamber suffered from often greater than 100\% error depending on where you sat. The visual observing labs suffered from Chapel Hill’s weather, bright skies, proximity to athletic field lights ruining dark adaptation, inability to see the north star, which is necessary to properly align the telescopes, outdated and difficult to use telescopes, and a weak set of backup labs.  Finally, neither set of labs strongly reinforced the lecture curriculum.  Feedback from these labs was generally negative.

We developed a series of eight new labs, two of which are two-week labs, and six of which utilize UNC-CH’s Skynet Robotic Telescope Network.  After an introductory lab in which students learn how to use Skynet, the labs strongly reinforce both the new ASTR 101/102 lecture curriculum and one another.  Among other things, students use Skynet to collect their own data to distinguish between geocentric and heliocentric models using the phase and angular size of Venus, to measure the mass of a Jovian planet using the orbit of one of its moons and Kepler’s third law, to measure the distance to an asteroid using parallax measured simultaneously by Skynet telescopes in different hemispheres, and to measure the distance to a globular cluster using an RR Lyrae star as a standard candle.  More is done with archival data that takes longer than a semester to collect (e.g., Cepheid stars, Type Ia supernovae, etc.)

In addition to the lecture, demo and lab curricula, we developed a set of multiple-choice homework problems and detailed solutions for both Astro 101 and Astro 102 within the WebAssign framework. Also, in an effort to explore the effectiveness of “flipping the classroom”, we developed a set of in-class polling questions, and an interactive e-polling tool that allows the instructor to display and analyze numerical responses in real-time. We also provided all students free online access via YouTube to a complete archive of videotaped Astro 101 and, soon, Astro 102 lectures compiled from previous semesters.

After implementing this curriculum at UNC-CH in 2009,  lab enrollments increased over 150\%, all introductory astronomy enrollments increased over 100\% – now one in four UNC-CH students take at least one of our courses – and astronomy-track majors and minors increased $\approx$300\% (from $\approx$5 to $\approx$20 per year). Encouraged by this initial success, we soon began partnering with other regional institutions to help them adopt and adapt those parts of the lecture course, in-class exercises and demos, homework, and labs that were compatible with their broader curricula and educational philosophies. As of today, 14 institutions have adopted our curriculum in whole or in part, with a handful more scheduled to join in the coming year. In this report, we analyze student survey responses collected from 10 schools, ranging from 2-year community colleges to Research I universities, over 4 semesters in 2014-2015.

While we provided instructors at these partner institutions access to our full sets of homework, lab, e-book, e-polling, video, and other curriculum resources, they were free to accept, reject or adapt any element to best suit their institutional needs and educational goals. Table 1 summarizes the institutions that employed our curriculum in whole or in part, and whose students responded to the end-of-course survey, during the period of 2014-2015. Table 2 describes in greater detail the components of our curriculum that each instructor chose to implement in their section.
\begin{table*}[hbt]
\caption{Summary of institutional participation, by institution. Institution types: 1 = 2-year community college; 2 = 4-year college or university; 3 = Research I university}
\centering
\begin{tabular}{|l|l|l|l|l|l|}
\hline
\textbf{Institution}                                                                                                & \textbf{Type} & \textbf{Semesters} & \textbf{Sections} & \textbf{Instructors} & \textbf{Responses} \\ \hline
\begin{tabular}[c]{@{}l@{}}Ashland Community \\ \& Technical College (ACTC)\end{tabular}                                                               & 2    & 2         & 3        & 1           & 12        \\ \hline
Francis Marion University (FMU)                                                                             & 2    & 1         & 1        & 1           & 6         \\ \hline
Fayetteville State University (FSU)                                                                         & 2    & 3         & 6        & 1           & 34        \\ \hline
Glenville State College (GSC)                                                                               & 2    & 1         & 1        & 1           & 5         \\ \hline
High Point University (HPU)                                                                                 & 2    & 2         & 2        & 1           & 15        \\ \hline
\begin{tabular}[c]{@{}l@{}}North Carolina Agricultural \& Technical \\ State University (NCAT)\end{tabular} & 2    & 3         & 3        & 2           & 29        \\ \hline
North Carolina State University (NCSU)                                                                      & 3    & 2         & 2        & 1           & 6         \\ \hline
\begin{tabular}[c]{@{}l@{}}University of North Carolina \\ at Chapel Hill (UNC-CH)\end{tabular}             & 3    & 4         & 11       & 2           & 427       \\ \hline
University of Virginia (UVa)                                                                                & 3    & 2         & 2        & 1           & 28        \\ \hline
\begin{tabular}[c]{@{}l@{}}Wake Technical \\ Community College (WTCC)\end{tabular}                                                                     & 1    & 4         & 10       & 5           & 187       \\ \hline
\textbf{Total=10}                                                                                                   & \textbf{NA}   & \textbf{4}         & \textbf{41}       & \textbf{16}          & \textbf{749}       \\ \hline
\end{tabular}
\end{table*}

\begin{table*}[hbt]
\caption{More detailed breakdown of student survey responses by section, including number of UNC labs and homeworks each instructor utilized, whether they used any of the Skynet-based telescope labs, whether the section was online, and whether any there was any lab component to the course at all. }
\centering
\begin{tabular}{|l|l|l|l|l|l|l|l|l|}
\hline
\textbf{Semest.} & \textbf{Institut.} & \textbf{Instructor} & \textbf{\# Resp.} & \textbf{UNC Labs} & \textbf{UNC HWs} & \textbf{Skynet} & \textbf{Online} & \textbf{Labs} \\ \hline
2014 S           & ACTC               & Riggs               & 7                 & 8                 & 0                & Y               & Y               & Y             \\ \hline
2014 S           & FSU                & Mattox              & 8                 & 1                 & 9                & N               & N               & Y             \\ \hline
2014 S           & FSU                & Mattox              & 4                 & 0                 & 9                & N               & N               & Y             \\ \hline
2014 S           & GSC                & O'Dell              & 5                 & 5                 & 0                & Y               & N               & Y             \\ \hline
2014 S           & HPU                & Barlow              & 10                & 2                 & 0                & Y               & N               & Y             \\ \hline
2014 S           & NCAT               & Schuft              & 16                & 0                 & 8                & N               & N               & N             \\ \hline
2014 S           & WTCC               & Chilton             & 18                & 0                 & 0                & N               & N               & Y             \\ \hline
2014 S           & WTCC               & Converse            & 22                & 0                 & 0                & N               & N               & N             \\ \hline
2014 S           & WTCC               & Wetli               & 11                & 5                 & 0                & Y               & N               & Y             \\ \hline
2014 S           & UNC                & Law                 & 46                & 8                 & 9                & Y               & N               & N             \\ \hline
2014 S           & UNC                & Reichart            & 5                 & 8                 & 9                & Y               & Y               & Y             \\ \hline
2014 S           & UNC                & Reichart            & 22                & 8                 & 9                & Y               & N               & Y             \\ \hline
2014 S           & UVA                & Murphy              & 24                & 0                 & 0                & N               & N               & Y             \\ \hline
2014 S           & ACTC               & Riggs               & 1                 & 8                 & 0                & Y               & Y               & N             \\ \hline
2014 F           & FSU                & Mattox              & 7                 & 1                 & 9                & N               & N               & N             \\ \hline
2014 F           & FSU                & Mattox              & 8                 & 0                 & 9                & N               & N               & Y             \\ \hline
2014 F           & FMU                & Bryngelson          & 6                 & 7                 & 0                & Y               & N               & Y             \\ \hline
2014 F           & NCAT               & Schuft              & 12                & 0                 & 5                & N               & N               & Y             \\ \hline
2014 F           & WTCC               & Converse            & 34                & 2                 & 0                & Y               & N               & Y             \\ \hline
2014 F           & WTCC               & Wetli               & 8                 & 4                 & 0                & Y               & N               & Y             \\ \hline
2014 F           & UNC                & Reichart            & 193               & 8                 & 9                & Y               & N               & Y             \\ \hline
2014 F           & UNC                & Reichart            & 11                & 8                 & 9                & Y               & Y               & Y             \\ \hline
2015 S           & UNC                & Reichart            & 4                 & 8                 & 9                & Y               & Y               & Y             \\ \hline
2015 S           & HPU                & Barlow              & 5                 & 2                 & 0                & Y               & N               & Y             \\ \hline
2015 S           & NCSU               & Frohlich            & 4                 & 8                 & 0                & Y               & N               & Y             \\ \hline
2015 S           & WTCC               & Converse            & 16                & 2                 & 0                & Y               & N               & Y             \\ \hline
2015 S           & WTCC               & Wetli               & 7                 & 4                 & 0                & Y               & N               & Y             \\ \hline
2015 S           & NCAT               & Kebede              & 1                 & 8                 & 9                & Y               & N               & Y             \\ \hline
2015 S           & UNC                & Reichart            & 21                & 8                 & 9                & Y               & N               & Y             \\ \hline
2015 S           & UVA                & Murphy              & 4                 & 7                 & 0                & Y               & N               & N             \\ \hline
2015 S           & WTCC               & Chilton             & 32                & 2                 & 0                & Y               & N               & Y             \\ \hline
2015 F           & ACTC               & Riggs               & 4                 & 8                 & 0                & Y               & N               & N             \\ \hline
2015 F           & FSU                & Mattox              & 5                 & 1                 & 9                & N               & N               & N             \\ \hline
2015 F           & FSU                & Mattox              & 2                 & 0                 & 9                & N               & N               & Y             \\ \hline
2015 F           & WTCC               & Converse            & 29                & 2                 & 0                & Y               & N               & Y             \\ \hline
2015 F           & NCSU               & Frohlich            & 2                 & 3                 & 0                & Y               & N               & Y             \\ \hline
2015 F           & UNC                & Reichart            & 109               & 8                 & 9                & Y               & N               & Y             \\ \hline
2015 F           & UNC                & Reichart            & 9                 & 8                 & 9                & Y               & Y               & Y             \\ \hline
2015 F           & UNC                & Reichart            & 7                 & 8                 & 9                & Y               & Y               & Y             \\ \hline
2015 F           & WTCC               & Sivayogan           & 10                & 4                 & 0                & Y               & N               & Y             \\ \hline
\end{tabular}
\end{table*}

\section*{Survey Structure, and Definition of Dependent and Independent Variables}

Near the end of each semester, students in participating sections were provided with a link to a Qualtrics survey about their experience in Introductory Astronomy. For the four semesters analyzed in this report, we received an initial total of 827 completed surveys. After eliminating incomplete or obviously fraudulent instances, we arrived at a final dataset of 749 responses.

The survey consists of 43 multiple-choice and short-answer questions, some of which consist of multiple parts. The questions include basic demographic information and assessments of a student’s background and preparation for the course, but are primarily geared towards determining a student’s opinion of the course and their attitude towards specific course components and towards astronomy and science in general. Some questions ask students to rank their opinion of a course component, or their level of agreement with a statement, on a four- or five-step scale (quantitative questions). A number of these quantitative survey questions consist of multiple sub-questions. A few questions are in yes/no format, or otherwise establish whether or not a student engaged with particular components of the course (binary questions). The full text of the survey can be downloaded at:
\href{https://tinyurl.com/introastroreport
}{https://tinyurl.com/introastroreport}

In order to facilitate analysis, responses to all questions were reassigned to a uniform numerical scale ranging from -1 to +1. For binary questions, this is as simple as assigning a “Yes” answer the value +1, and a “No” answer the value -1. For quantitative questions, this required both renormalizing the numerical range of the responses, and, in some cases, flipping the sign of the response to correct for whether the question had a “positive” or “negative” attitudinal orientation. 

The responses to some multi-part quantitative questions were averaged (after numerical range normalization and attitudinal orientation correction) to produce a single numerical index for that question. An illustrative example is the astronomy/science “Attitude Index”, which serves as the single dependent variable in the analysis that follows. This Attitude Index is computed from the respondents’ answers to 33 questions that were designed to probe their attitudes towards Astronomy and science in general, after having taken Introductory Astronomy at their institution. Each question is in the form of a statement; students were instructed to indicate their level of agreement with each statement, from 1 (strongly disagree) to 3 (neither agree nor disagree) to 5 (strongly agree). By design, some statements were positively oriented (e.g., “I like astronomy”, “Scientific concepts are easy to understand”, “Scientific skills will make me more employable”), while some were negatively oriented (e.g., “Astronomy is irrelevant to my life”, “I felt insecure when I had to do astronomy homework”, “I find it difficult to understand scientific concepts”). Each response was converted to a numerical scale ranging from -1 (negative attitude) to +1 (positive attitude), taking into account the orientation of each question, and the results were averaged over the 33 questions, producing a single Attitude Index for each student respondent. While the perceived orientation of certain of these statements may be qualitative, with different students seeing the same statement as either positive or negative, the majority are unambiguous. The orientations we assigned to the Attitude Index questions are presented in Table 3.

\begin{table*}[hbt]
\caption{The questions that were used to compute the astronomy and science Attitude Index dependent variable. Student responses to each statement were scaled from -1 = strongly disagree to +1 = strongly agree. The sign of responses was flipped for those statements with a science-negative orientation, and then all were averaged to arrive at the Attitude Index.}
\begin{tabular}{|l|l|}
\hline
\textbf{\begin{tabular}[c]{@{}l@{}}Attitude Index   Question\end{tabular}} & \textbf{Orientation} \\ \hline
Astronomy is a subject learned quickly by most people.                        & +                    \\ \hline
I have trouble understanding astronomy because of how I think.                & -                    \\ \hline
Astronomy concepts are easy to understand.                                    & +                    \\ \hline
Astronomy is irrelevant to my life.                                           & -                    \\ \hline
I was under stress during astronomy class.                                    & -                    \\ \hline
I understand how to apply analytical reasoning to astronomy.                  & +                    \\ \hline
Learning astronomy requires a great deal of discipline.                       & -                    \\ \hline
I have no idea of what's going on in astronomy.                               & -                    \\ \hline
I like astronomy.                                                             & +                    \\ \hline
What I learned in astronomy will not be useful in my career.                  & -                    \\ \hline
Most people have to learn a new way of thinking to do astronomy.              & -                    \\ \hline
Astronomy is highly technical.                                                & -                    \\ \hline
I felt insecure when I had to do astronomy homework.                          & -                    \\ \hline
I find it difficult to understand astronomy concepts.                         & -                    \\ \hline
I enjoyed taking this astronomy course.                                       & +                    \\ \hline
I made a lot of errors applying concepts in astronomy.                        & -                    \\ \hline
Astronomy involves memorizing a massive collection of facts.                  & -                    \\ \hline
Astronomy is a complicated subject.                                           & -                    \\ \hline
I can learn astronomy.                                                        & +                    \\ \hline
Astronomy is worthless.                                                       & -                    \\ \hline
I am scared of astronomy.                                                     & -                    \\ \hline
Science is a part of everyday life.                                           & +                    \\ \hline
Scientific concepts are easy to understand.                                   & +                    \\ \hline
Science is not useful to the typical professional.                            & -                    \\ \hline
The thought of taking a science course scares me.                             & -                    \\ \hline
I like science.                                                               & +                    \\ \hline
I find it difficult to understand scientific concepts.                        & -                    \\ \hline
I can learn science.                                                          & +                    \\ \hline
Scientific skills will make me more employable.                               & +                    \\ \hline
Science is a complicated subject.                                             & -                    \\ \hline
I use science in my everyday life.                                            & +                    \\ \hline
Scientific thinking is not applicable to my life outside my job.              & -                    \\ \hline
Science should be a required part of my professional training.                & +                    \\ \hline
\end{tabular}
\end{table*}

In the analysis that follows, we explore the statistical dependence of Attitude Index (dependent variable) on a variety of other survey responses/indices (independent variables), using simultaneous multiple linear regression. After initially performing linear regression with 16 independent variables, we iteratively removed those independent variables that were uncorrelated with Attitude Index at the $p > 0.05$ level, refitting at each iteration. The results are summarized in Table 4. We found the following variables to exhibit \textbf{significant correlation} (in decreasing order of correlation coefficient):
\begin{itemize}[noitemsep] 

\item	\textbf{Course Attitude Index (Q48 in original survey; see Appendix):} measures a student’s attitude to the course as a whole, based on an average of responses to 10 statements, scaled to -1 = strongly disagree to +1 = strongly agree. \textbf{Positively correlated.}
\item	\textbf{Grade Index (Q17):} what grade students \textit{expected to receive} in the course at the time they took the survey. -1 = F, 0 = C, +1 = A. \textbf{Positively correlated.}
\item	\textbf{Career Index (Q12):} measures the degree to which a student’s academic and career path is oriented towards STEM in general, and astronomy \& physics in particular. -1 = planning a career in a non-STEM field; 0 = planning a career in a STEM field; 1 = Planning a career in a STEM field, and majoring or minoring in astronomy or physics. \textbf{Positively correlated.}
\item	\textbf{Instructor Index (Q64):} measures a student’s attitude towards the primary course instructor, based on an average of responses to 11 statements (Q64), scaled to -1 = strongly disagree to +1 = strongly agree. \textbf{Positively correlated.}
\item	\textbf{Math Index (Q8):} measures a student’s academic mathematics training background. Ranges from -1 = some algebra to +1 = beyond calculus. \textbf{Positively correlated.}
\item	\textbf{Institution Index (see Table 1):} measure of the type of institution the course was offered at: -1 = 2yr college, 0 = 4yr college, +1 = research I university. \textbf{Negatively correlated.}
\item	\textbf{Work Index (Q23):} based on the response to the statement ``I worked harder than I thought I would in order to meet the instructor's standards or expectations.'' -1 = strongly disagree to +1 = strongly agree. \textbf{Negatively correlated.}
\end{itemize}

The following independent variables were found to exhibit \textbf{no significant correlation} with Attitude Index at the $p < 0.05$ level:

\begin{itemize}[noitemsep]
\item	\textbf{Skynet Index (see Table 2):} -1 = student was offered no Skynet-based labs; +1 = student was offered Skynet-based labs.
\item	\textbf{Lab Index (see Table 2):} -1 = no lab component to course at all; +1 = some lab component to course.
\item	\textbf{Online Index (see Table 2):} -1 = traditional lecture course; +1 = online course.
\item	\textbf{Engagement Index (Q35):} measures a student’s level of engagement with the course, based on an average of their responses to 6 questions about how often they employed various study habits (doing readings, completing assignments, engaging in classroom discussion, etc.).
\item	\textbf{Hours Index (Q15):} the number of hours the student spent per week on course-related work. Ranges from -1 = fewer than 3, to 0 = 7-9 hours, to +1 = 12 or more hours.
\item	\textbf{Credits Index (Q13):} how many credit hours the student was enrolled in while taking the intro astro course. Ranges from -1 = 6 or fewer credit hours to 0 = 7-9 credit hours to +1 = 19 or more credit hours.
\item	\textbf{Year Index (Q19):} the academic year of the student. Ranges from  -1 = first year to +1 = 5th+ year.
\item	\textbf{Attendance Index (Q22):} based on the question ``It is possible to do well in this course without attending class regularly'', ranges from -1 = strongly disagree to +1 = strongly agree.
\item	\textbf{UNC HW Index (see Table 2):} measure of how many UNC-provided homework sets were assigned in the student’s section. Ranges from -1 = none to +1 = all of the 9 available sets.
\end{itemize}

\section*{Baseline Model}
As described above, we found that 7 of our independent variables were significantly correlated with the Attitude Index at the $p < 0.05$ level; the results are summarized in Table 4.

\begin{table*}[hbt]
\caption{Multiple linear regression correlation coefficients for the entire survey data set ($N = 749$), after iterative elimination of independent variables for which $p > 0.05$.}
\begin{tabular}{|l|l|l|}
\hline
\textbf{Variable}         & \textbf{Coefficient} & \textbf{\textit{p}-value} \\ \hline
Course Attitude Index     & 0.25                 & 2.5E-26          \\ \hline
Grade Index               & 0.16                 & 1.6E-12          \\ \hline
Career Index              & 0.11                 & 2.0E-21          \\ \hline
Instructor Attitude Index & 0.073                & 5.4E-03          \\ \hline
Math Index                & 0.053                & 1.5E-04          \\ \hline
Institution Index         & -0.048               & 6.1E-06          \\ \hline
Work Index                & -0.10                & 7.8E-10          \\ \hline
\end{tabular}
\end{table*}

We consider each of these variables in turn, in descending order of correlation coefficient:
\begin{enumerate}[noitemsep] 
\item	\textbf{Course Attitude Index:} It is not surprising that the Astronomy/Science Attitude Index’s strongest and most significant correlation is that with the student’s attitude towards and opinion about the course overall. The questions that comprise the Course Attitude index (Q48 in survey) focus on whether a student feels that the course and the work involved were effective in helping them learn, whether sufficient feedback was provided on a student’s progress, and whether the student found the course inspiring and challenging. As with all of these correlations, we must speculate on causal relationships with caution. Does a positive experience in the course create a positive attitude towards science, or are students who were predisposed to view science favorably more likely to appreciate a course in introductory astronomy in the first place? It’s not possible to disentangle these two with this analysis, but we can at least infer that the most impactful strategy for an institution to take, if its goal is to increase positive attitudes towards science in general, is to foster positive attitudes towards the student experience of an introductory course itself -- its goals, pacing, feedback, and level of intellectual challenge. 
\item	\textbf{Grade Index:} It is also not surprising that a student’s attitude towards science in general, after taking an introductory science course, would be correlated with the grade that they expect to receive. As with the previous index, it is not possible to say whether this is just correlation or causation. But by accounting for these strongly correlated Grade and Course Attitude Indices in the simultaneous multiple linear regression analysis, we can at least begin to unmask some of the subtler correlations that follow. We chose to explore the self-reported expected grade both because it is much easier, logistically and ethically, than attempting to assign actual grades to ostensibly anonymous surveys, and because, when it comes to attitudes, a student’s self-perceived grade at the time of the survey is more likely to matter than what they actually end up getting.
\item	\textbf{Career Index:} After Course Attitude Index, the STEM Career Index is the most significantly correlated independent variable. Again, as with the previous variables, it is not possible to say whether students who had already decided to pursue STEM careers are predisposed to have more positive attitudes towards science, or whether positive attitudes engendered by the course prompted some students to consider STEM careers for the first time. This is a case where giving the survey both at the beginning and at the end of the course would be very helpful in interpreting the results. It is worth noting that 370 out of 749, or nearly 50\% of the total respondents indicated that they did not intend to pursue STEM-related careers. As a group, these non-STEM students receive a less positive impact on science attitude than do their general STEM-major peers, who in turn are impacted less than those who specifically plan careers in astronomy or physics. 
\item	\textbf{Instructor Index:} While it makes sense that students who view their instructor positively might emerge from the course with a more positive attitude towards science, it is interesting that the correlation, while positive, is both relatively low and marginally significant. Also, as discussed in the following section, when we look only at the subsets of students who attended a lecture course, or who were exposed to Skynet during the course, and include their ratings of these components’ helpfulness as independent variables in the regression analysis, the correlation of Attitude Index with Instructor Index disappears. The message seems to be that instructor quality helps to shape attitudes towards science, but not nearly as much as the perceived quality of the course curriculum and experience as a whole
\item	\textbf{Math Index:} This is another correlation that is unsurprisingly positive but surprisingly weak. Having a more extensive mathematical coursework background corresponds to more positive science attitudes at the end of the course, but not by much.  This would suggest that our Intro Astro curriculum (which requires only basic algebra) is relatively equally accessible and impactful to every student, regardless of mathematical background.
\item	\textbf{Institution Index:} There is a weak but statistically significant negative correlation of Attitude Index with the type of institution the course is offered at, with 2-year community colleges doing better, in general, than 4-year colleges, and both doing better than Research I universities. This trend may reflect a dependence on class size and instructor availability, with the smaller, more personal environments typical of community college classrooms serving better to instill positive attitudes towards science than large auditorium-style lecture formats typical of major universities.
\item	\textbf{Work Index:} Students who perceive the course to require more work than average emerge with more negative attitudes towards science. This would suggest that one straightforward way to boost science attitudes would be to reduce the workload in introductory science courses. It is important to keep in mind, however, that the very significantly positively correlated Course Attitude Index is partially a measure of how intellectually challenging and instructive the course is perceived to be. Make the course too easy, and you risk negatively impacting attitude towards the course, and so towards science in general. 
\end{enumerate}

\section*{Baseline Model + Helpfulness of Course Components}

Students were asked to rate the Helpfulness Index of various components of their introductory astronomy courses, if present, which we scale from -1 = not helpful to +1 = extremely helpful (Q52). The students were given the option to indicate that any component was not applicable to their experience. The course components included:

\begin{itemize}[noitemsep]
\item	Attending class lectures
\item	Watching videos of lectures
\item	Supplementary notes (e.g., e-book on \\ WebAssign)
\item	Homeworks
\item	In-class exercises/polling (e.g., clickers, polling cards, e-polling)
\item	Textbook
\item	Out-of-class exercises
\item	Office hours
\item	Online discussion forum (e.g., Sakai or Blackboard)
\item	Skynet-based telescope labs
\item	Other telescope labs (not part of UNC’s curriculum, but employed at some participating institutions)
\item	Non-telescope labs (e.g., The Earth and the Seasons, or Hubble’s Law)

\end{itemize}

We found that the mere \textit{presence} of any of these individual course components (included in the linear regression analysis as binary independent variables where -1 = not used in course and +1 = used in course) did not significantly impact the students’ attitudes about science. However, we did find that, for two components – attending in-class lectures, and Skynet-based labs – \textit{how helpful} the students found these course components to be did matter.  For the other course components, neither the existence of the component nor how helpful the students found it to be impacted the Attitude Index.

For each component, we analyzed the subset of non-N/A respondents and performed multiple linear regression on Attitude Index vs the same set of independent variables described earlier, but including the Helpfulness Index of that component, again iteratively eliminating independent variables for which the correlation significance was low. The Helpfulness Indices range from -1 = not helpful to my learning to +1 = extremely helpful to my learning.

\begin{figure}[ht]\centering 
\includegraphics[width=\linewidth]{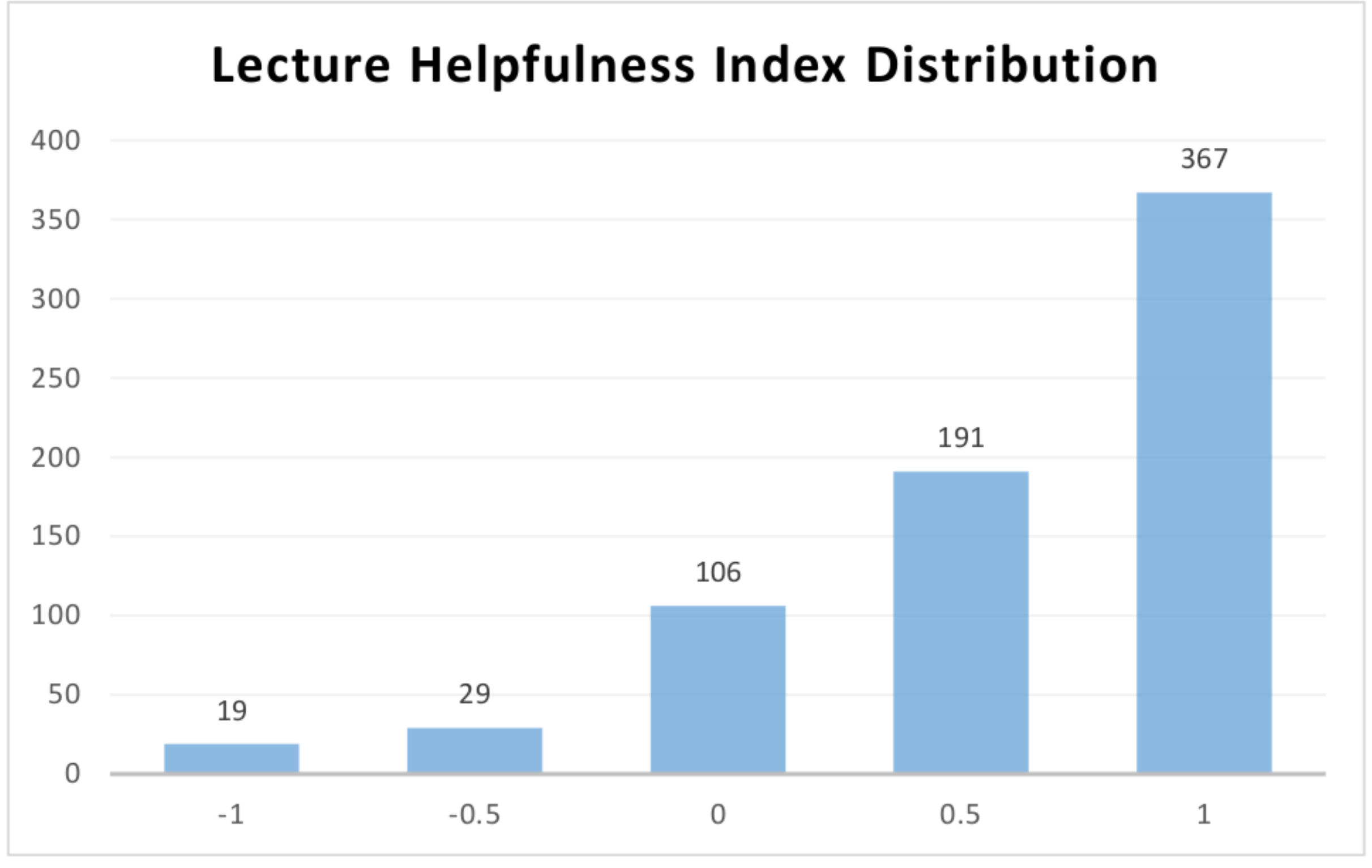}
\caption{Distribution of our sample of 712 students who rated the helpfulness of in-class lectures. Those
who found the lectures helpful left the course with more positive attitudes about astronomy and STEM fields in general. Other
than our Skynet-based labs, no other course component had a similar effect. The Helpfulness Indices
range from -1 = not helpful to my learning to +1 = extremely helpful to my learning.}
\label{fig:figure1}
\end{figure}

Out of the 749 total student responses to the survey, 712 students attended in-classroom lectures. The distribution of Lecture Helpfulness Index ratings in this subsample is plotted in Figure~\ref{fig:figure1},The results of multiple linear regression on this subset are presented in Table~5. We find a weak but statistically significant positive correlation between the Lecture Helpfulness Index and the Attitude Index: the more helpful a student finds attending class to be, the more positive their attitude towards science at the end of the course. However, note that the Instructor Attitude Index, which exhibited weak but significant positive correlation in the earlier baseline analysis (Table~4), is no longer significantly correlated in this subset ($p = 0.2$), when Lecture Helpfulness Index is included. Both are somewhat weak correlations, but this result would seem to indicate that students’ attitudes towards their instructors and towards the effectiveness of the lectures are largely measures of the same thing. Instructors who wish to improve their students’ attitudes towards science would thus be well served by investing more effort into polishing the lecture component of their course.

\begin{table*}[hbt]
\caption{Results of multiple linear regression on Attitude Index vs. significant variables, including Lecture Helpfulness Index, for the subset of $N = 712$ students who attended in-class lectures. \textit{Top}: Fit with Instructor Attitude Index (which is not correlated with Attitude Index at the $p<0.05$ level); \textit{Bottom}: Fit with Instructor Attitude Index excluded.}
\begin{tabular}{|l|l|l|}
\hline
\textbf{Variable}         & \textbf{Coefficient} & \textbf{p-value} \\ \hline
Course Attitude Index     & 0.24                 & 1.3E-21          \\ \hline
Grade Index               & 0.14                 & 4.2E-10          \\ \hline
STEM Index                & 0.11                 & 1.6E-18          \\ \hline
Lecture Helpfulness Index             & 0.071                & 1.7E-04          \\ \hline
Math Index                & 0.063                & 1.0E-05          \\ \hline
\textit{Instructor Attitude Index} & \textit{0.036}                & \textit{2.0E-01}          \\ \hline
Institution Index         & -0.048               & 6.6E-06          \\ \hline
Work Index                & -0.11                & 1.0E-09          \\ \hline
\multicolumn{3}{}{} \\ \hline
\textbf{Variable}     & \textbf{Coefficient} & \textbf{p-value} \\ \hline
Course Attitude Index & 0.26                 & 2.2E-28          \\ \hline
Grade Index           & 0.14                 & 2.5E-10          \\ \hline
STEM Index            & 0.11                 & 2.6E-18          \\ \hline
Lecture Helpfulness Index         & 0.077                & 2.1E-05          \\ \hline
Math Index            & 0.064                & 7.7E-06          \\ \hline
Institution Index     & -0.046               & 1.2E-05          \\ \hline
Work Index            & -0.11                & 1.0E-09          \\ \hline
\end{tabular}
\end{table*}

\begin{figure}[ht]\centering 
\includegraphics[width=\linewidth]{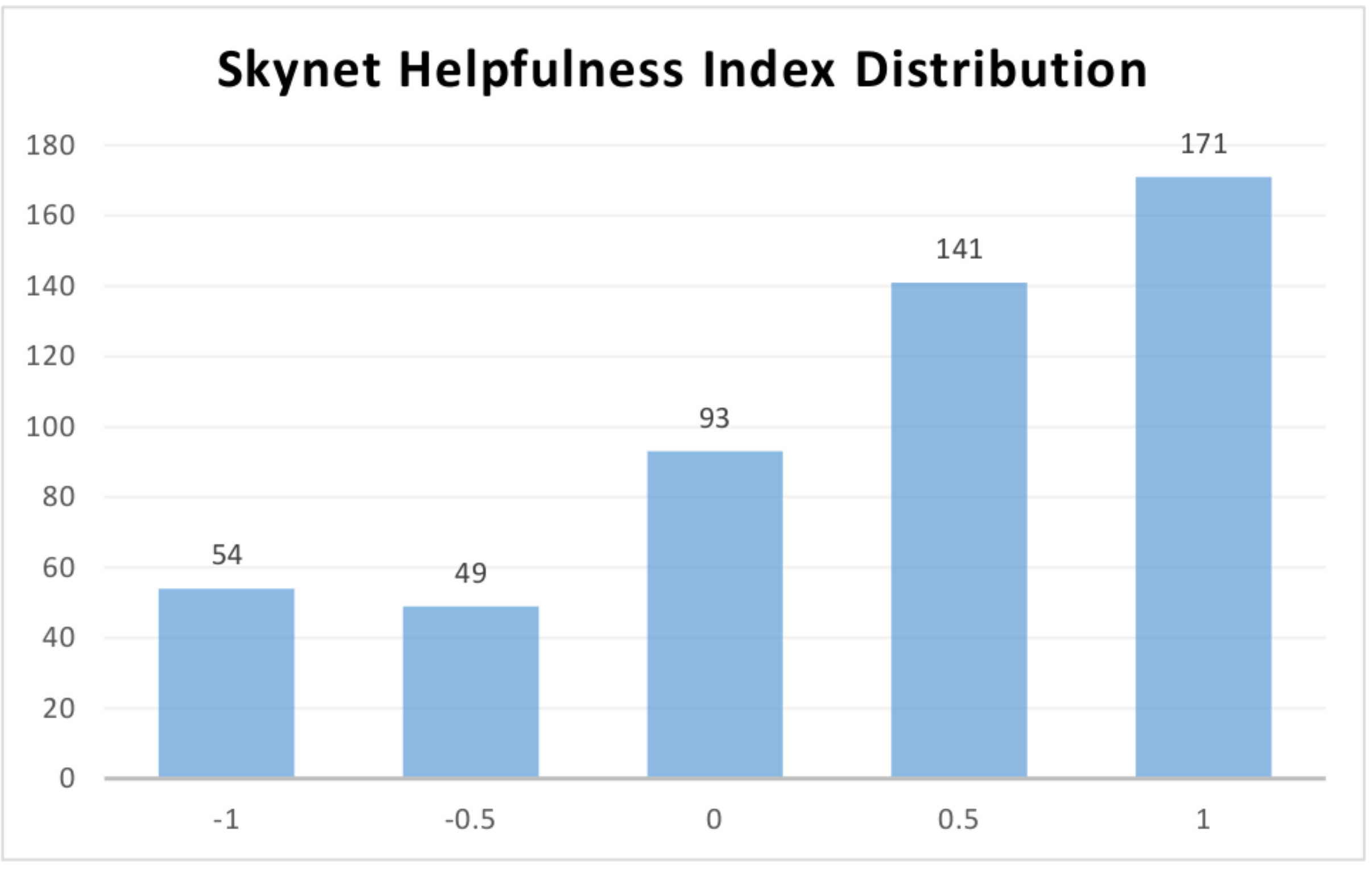}
\caption{Distribution of our sample of 508 students who rated the helpfulness of Skynet-based telescope labs. Those who rated these labs as helpful left the course with more positive attitudes about astronomy and STEM fields
in general. Other than in-class lectures, no other course component had a similar effect. The
Helpfulness Indices range from -1 = not helpful to my learning to +1 = extremely helpful to my learning.}
\label{fig:figure2}
\end{figure}

Out of the 749 total student responses to the survey, 508 students participated in at least one Skynet-based lab during the semester. The distribution of Skynet Helpfulness Index ratings in this subsample is plotted in Figure~\ref{fig:figure2}, and the results of multiple linear regression on this subset are presented in Table~6. As with Lecture Helpfulness, we find a weak but statistically significant correlation between Science Attitude Index and Skynet Helpfulness Index for the subset of students who were exposed to Skynet-based labs. Those students who found the labs helpful, left the course with a more positive attitude towards science overall. Note that we performed this same analysis for the helpfulness of other lab components, including indoor labs and non-Skynet-based telescope labs, and found no significant impact. Adding at least one Skynet-based lab (and working to present it and integrate it in a way that is perceived as helpful to students’ understanding of the course material) appears to be one way to significantly boost attitudes towards science for Intro Astro students.

\begin{table*}[hbt]
\caption{Results of multiple linear regression on Attitude Index vs. significant variables, including Skynet Helpfulness Index, for the subset of $N = 508$ students who participated in Skynet-based telescope labs. \textit{Top}: Fit with Instructor Attitude Index (which is not correlated with Attitude Index at the $p<0.05$ level); \textit{Bottom}: Fit with Instructor Attitude Index excluded.}
\begin{tabular}{|l|l|l|}
\hline
\textbf{Variable}         & \textbf{Coefficient} & \textbf{p-value} \\ \hline
Grade Index               & 0.21                 & 4.1E-13          \\ \hline
Course Attitude Index     & 0.20                 & 5.8E-11          \\ \hline
Career Index              & 0.11                 & 1.1E-13          \\ \hline
\textit{Instructor Attitude Index} & \textit{0.057}                & \textit{6.5E-02}          \\ \hline
Skynet Helpfulness Index              & 0.044                & 1.5E-02          \\ \hline
Math Index                & 0.036                & 5.1E-02          \\ \hline
Institution Index         & -0.050               & 1.2E-04          \\ \hline
Work Index                & -0.095               & 1.2E-05          \\ \hline
\multicolumn{3}{c}{} \\
\hline
\textbf{Variable}     & \textbf{Coefficient} & \textbf{p-value} \\ \hline
Course Attitude Index & 0.22                 & 1.4E-14          \\ \hline
Grade Index           & 0.22                 & 1.1E-13          \\ \hline
Career Index          & 0.11                 & 1.2E-13          \\ \hline
Skynet Helpfulness Index          & 0.047                & 8.8E-03          \\ \hline
Math Index            & 0.037                & 4.6E-02          \\ \hline
Institution Index     & -0.047               & 2.7E-04          \\ \hline
Work Index            & -0.095               & 1.4E-05          \\ \hline
\end{tabular}
\end{table*}

\section*{Conclusion}

The majority of students do not enroll in Introductory Astronomy as part of their major; for many, it is the last science course they will ever take, and has the potential to shape their attitudes towards STEM fields for the rest of their life. It is less important, therefore, when assessing the effectiveness of Intro Astro courses to explore traditional curricular learning gains, than it is to explore the effects that various course components have on this attitude. We first arrived at a baseline model (Table 5) describing the correlation, for the entire sample, of Attitude Index with a variety of independent variables describing students’ attitudes, backgrounds, and plans. We then analyzed, one at a time, subsets of the sample that reported engaging with various course components, and included as a new independent variable their rating of each component’s helpfulness. 

We found that the only course components whose helpfulness indices exhibit correlation with overall astronomy and STEM attitudes were in-class lectures and Skynet-based labs. While considerable effort has been expended to add new components to the Intro Astro curriculum, from in-class e-polling systems and questions, to providing videotapes of lectures to all students, to writing supplementary e-book materials, we cannot say at this time that they have had any effect one war or the other on attitudes. That is not to say that they have no effects at all – they very well may be found to improve traditional learning outcomes, for instance. But the results of this analysis suggest that an instructor’s best bet for boosting attitudes with our Intro Astro curriculum is to concentrate on improving the quality of their lectures and of the Skynet-based telescope labs that they offer.

\phantomsection
\section*{Acknowledgments} 

\addcontentsline{toc}{section}{Acknowledgments} 

We gratefully acknowledge the support of the National Science Foundation, through the following programs and awards:  ESP 0943305, MRI-R2 0959447, AAG 1009052, 1211782, and 1517030, ISE 1223235, HBCU-UP 1238809, TUES 1245383, and STEM+C 1640131.  We are also appreciative to have been supported by the Mt. Cuba Astronomical Foundation, the Robert Martin Ayers Sciences Fund, and the North Carolina Space Grant Consortium.  The authors wish to thank Collin Wallace and Michael Fitzgerald for their very helpful and thoughtful comments on this work.

\phantomsection

\bibliography{biblio}


\end{document}